\documentclass[11pt,a4paper]{article}

\usepackage{amsfonts}
\usepackage[dvips]{graphicx}
\usepackage{amsthm}
\usepackage{amsmath}
\usepackage{epsfig}
\usepackage{amssymb}
\usepackage{latexsym}
\usepackage{bm}

\begin{document}

\renewcommand{\figurename}{Fig.}

\centerline {\LARGE \bf Spinning particle in a varying magnetic
field:}

\vskip 1mm \centerline {\LARGE \bf how work is done by changing
external parameter}

\vskip 5mm \centerline{\large Vladim\' ir Balek\footnote{e-mail
address: balek@fmph.uniba.sk}}

\vskip 5mm \centerline{\large \it Department of Theoretical
Physics, Comenius University, Bratislava, Slovakia}

\vskip 1cm

{\small The work done by a spinning particle, or on it, when put
into a varying magnetic field is discussed.}

\vskip 5mm

If a spinning particle with magnetic moment $\mu$ is located in a
magnetic field that increases from zero to $B$ and points in the
direction of the moment, the particle loses energy $\mu B$. If the
field behaves the same way but points in the opposite direction,
the particle gains energy $\mu B$. The particle performs work $\mu
B$ on the field in the former case and the field performs work
$\mu B$ on the particle in the latter case. Clearly, `to perform
work on the field' means to perform it on the equipment generating
the field, and `to perform work on the particle' means to perform
it on movable parts of the equipment against the action of the
magnetic field of the particle. In a similar way as `perform work
on a spinning particle' we can `perform work on a weight', if we
press the spring to which the weight is attached without moving
the weight itself. Work is defined as force times the distance
traveled by the object on which the force is acting, hence it can
be performed on a standing object only in a figurative sense; in
actual fact it is performed on an intermediate system attached to
the object (spring, equipment generating the field).

\vskip 2mm Work is supplied to the field or done by it also if the
field is acting on a system of spinning particles in contact with
heat reservoir. In this setting, the field can be viewed as an
external parameter and the work can be computed from the standard
formula `pressure times the increment of the parameter'. The
`pressure', however, is not the force per unit area coming from
the chaotic motion of the particles, but the component of the
total magnetic moment in the direction of the field. As it turns
out, this quantity is related to the magnetic field in the same
way as the ordinary pressure is related to the volume.

\vskip 2mm The identification of magnetic field with an external
parameter, and the resulting formula for work, are well known to
any student of thermodynamics. Magnetic field is mentioned on
regular basis as {\it the} second example of an external
parameter, the first being volume. However, to see how the work
defined by the thermodynamic formula can actually be extracted
from the system, that is to say, how a string of spins in a
varying magnetic field can be used 'to turn the wheel' somewhere
in the equipment generating the field, we must go back to the
description of magnetic field in Maxwell theory.

\vskip 2mm {\it 1. Thermodynamics of a string of spins.} Consider
$N$ particles with spin 1/2 and unit magnetic moment, put into
homogeneous magnetic field $B$. The spins are supposed to have
only two orientations, in the direction of the field (upwards) or
in the direction opposite to the field (downwards). The energy of
the system is
\begin{equation}
E = - MB, \label{eq:en}
\end{equation}
with the total magnetic momentum $M$ given by
\begin{equation}
M = 2n - N. \label{eq:mo}
\end{equation}
where $n$ is the number of spins oriented upwards. The entropy of
the system is
$$S = \log \mbox{\Large $\left({}^N \mbox{\hskip -3mm ${}_n$}
\right)$}.$$ (We use a system of units in which $k = 1$.) For
large $N$, $n$ and $N - n$ it holds
$$\mbox{\Large $\left({}^N \mbox{\hskip -3mm ${}_n$} \right)$} \simeq
\frac {N^N}{\sqrt{2\pi} n^n (N - n)^{N - n}},$$
thus
\begin{equation}
S \simeq N \log N - n \log n - (N - n) \log (N - n).
\label{eq:ent}
\end{equation}
We have omitted the constant $- \log \sqrt{2\pi}$ in the
expression for $S$ since it has no effect on the results. Note
that without it, $S$ assumes the correct value $S = 0$ for $n = 0$
and $N$. The inverse temperature can be computed from
$$\beta = \partial_E S = \frac 1{2B} \partial_n S.$$
By inserting here the function $S(n)$ from (\ref{eq:ent}), we find
\begin{equation}
\beta = \frac 1{2B} \log \frac {N - n}n. \label{eq:be}
\end{equation}
Note that we obtain the same formula when computing $\partial_n S$
as $\displaystyle \log \mbox{\Large $\left( \mbox{\hskip 2mm
${}^N$} \mbox{\hskip -5mm ${}_{n + 1}$} \right)$} - \log
\mbox{\Large $\left({}^N \mbox{\hskip -3mm ${}_n$} \right)$}$;
however, we arrive at a {\it wrong} formula when using an
apparently equivalent expression $\displaystyle \mbox{\Large
$\left({}^N \mbox{\hskip -3mm ${}_n$} \right)$}^{-1} \left[
\mbox{\Large $\left( \mbox{\hskip 2mm ${}^N$} \mbox{\hskip -5mm
${}_{n + 1}$} \right)$} - \mbox{\Large $\left({}^N \mbox{\hskip
-3mm ${}_n$} \right)$} \right]$, the reason being that one has to
put $\Delta n \ll 1$ rather than $\Delta n = 1$ when computing the
derivative $\displaystyle \partial_n \mbox{\Large $\left({}^N
\mbox{\hskip -3mm ${}_n$} \right)$}$ as the ratio of increments.
(Two errors cancel each other!) From (\ref{eq:be}) we can
immediately see that the temperature is positive for $n < N/2$ and
negative for $n > N/2$. Having found $\beta$ as a function of $n$,
we can express $n$ and, by using equations (\ref{eq:mo}) and
(\ref{eq:ent}), $M$ and $S$, as functions of $\beta$. In this way
we obtain
\begin{equation}
M = N \tanh \beta B \label{eq:mo1}
\end{equation}
and, after some algebra,
\begin{equation}
S = - \beta MB + N \log (2 \cosh \beta B). \label{eq:ent1}
\end{equation}
The second formula can be checked by computing $S$ from the
thermodynamic definition,
$$S = \int \beta \delta Q,$$
where $\delta Q$ is the heat received by the system and the
integral is to be taken over a reversible process. If $B = const$,
work is done neither by, nor on, the string of spins, therefore
the heat received in an arbitrary process is $\delta Q = dE$. If
we make use of (\ref{eq:en}) with $B = const$, we obtain
\begin{equation}
\delta Q = - B dM. \label{eq:he}
\end{equation}
This, together with the expression (\ref{eq:mo1}) for $M$ (valid
for systems in thermodynamic equilibrium, hence applicable to
reversible processes), yields again the expression (\ref{eq:ent1})
for $S$.

\vskip 2mm The `pressure' corresponding to the magnetic field is
$$p_B \equiv - (\partial_B E)_S = M,$$
where we have used the fact that $S$ can be expressed in terms of
$M$ only, see equations (\ref{eq:mo}) and (\ref{eq:ent}). The work
done by the string of spins when the magnetic field increases by
$dB$ is $\delta A = p_B dB$, or
\begin{equation}
\delta A = MdB. \label{eq:wo}
\end{equation}
From (\ref{eq:en}) we have $dE = - MdB - BdM$, hence the
expression (\ref{eq:he}) for $\delta Q$ and the recently obtained
expression for $\delta A$ sum up to the first law of
thermodynamics,
$$dE =  - \delta A + \delta Q.$$

\begin{figure}[ht]
\centerline{\includegraphics[width=14cm]{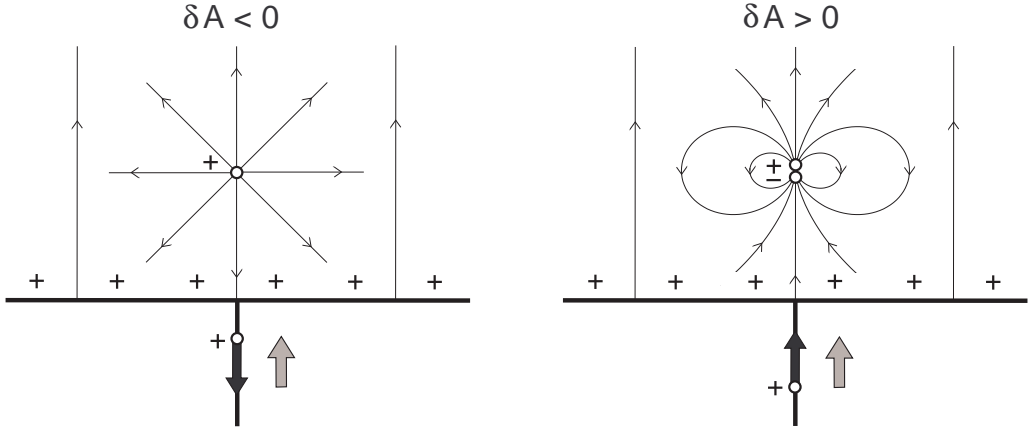}}
\caption{Charges in a varying electric field} \label{fig:eld}
\end{figure}

\vskip 2mm {\it 2. Electric charge in a varying electric field.}
Consider a pointlike charge $q$ in an external electric field,
say, placed nearby a charged metal plate (fig. \ref{fig:eld} to
the left). Suppose
an additional charge $\delta Q$ is brought onto the plate from
infinity, and denote the corresponding variation of the potential
$\Phi$ at the point where the charge is located by $\delta \Phi$.
The total work needed for this variation equals the work $\delta
A_0$ done by the external forces that compensate the Coulomb force
of the charges located on the plate, minus the work $\delta A$
done by the Coulomb forces of the charge $q$. The latter quantity
can be viewed as the net energy that can be extracted from the
process thanks to the fact that the charge $q$ is participating in
it. Note, however, that this can be understood literally only if
the positions of the charges producing the external field are not
influenced by the presence of the charge $q$. If they are, as is
the case for a metal plate, the work $\delta A_0'$ needed to
change the potential by $\delta \Phi$ when the plate is left on
its own, differs in general from the work $\delta A_0$ needed for
the same purpose when the charge $q$ is there, and the net energy
gain is $\delta A_0 + \delta A - \delta A_0'$ rather than $\delta
A$. Keeping in mind this reservation, let us find how $\delta A$
is related to $\delta \Phi$. Let $dQ$ be a portion of the charge
$\delta Q$ transported along the path ${\bf r} = {\bf r}(t)$ from
the point $Q_0$ at infinity to the point $Q$ on the plate. The
charge $q$ acts on the charge $dQ$ by the Coulomb force $d{\bf
f}_C = dQ {\bm \epsilon}$, where ${\bm \epsilon}$ is the intensity
of the electric field of the charge $q$ along the trajectory of
the charge $dQ$. Thus, the work done by the charge $q$ on the
charge $dQ$ is
$$dA = dQ \int \limits_{Q_0}^Q {\bm \epsilon}\ .\ d{\bf r}.$$
The force $d{\bf f}_C$ equals, by the law of action and reaction,
minus the Coulomb force $d{\bf f}_C'$ by which the charge $dQ$
acts on the charge $q$; that is, $d{\bf f}_C = - q \vec{\cal
E}_{dQ}$, where $\vec{\cal E}_{dQ}$ is the intensity of the
electric field of the charge $dQ$ at the point where the charge
$q$ is located. Let us pass from $d{\bf f}_C$ to $d{\bf f}_C'$,
and at the same time from the path ${\bf r} = {\bf r}(t)$ of the
charge $dQ$, with the charge $q$ fixed at the point $P$, to the
path ${\bf r}' = {\bf r}_P + {\bf r}_Q - {\bf r}(t)$ of the charge
$q$, with the charge $dQ$ fixed at the point $Q$. In this way we
obtain
$$dA = q \int \limits_{P_0}^P \vec{\cal E}_{dQ}\ .\ d{\bf
r},$$ where the point $P_0$ is located at infinity symmetrically
to the the point $Q_0$ with respect to the center of the segment
$PQ$. The rewritten expression for $dA$ reduces to
$$dA = - q \Phi_{dQ}(P, Q),$$
where $\Phi_{dQ}(P, Q)$ is the potential generated at the point
$P$ by the charge $dQ$ located at the point $Q$. Now, let $dQ$ be
a charge that has been located on the plate from the very start,
and was shifted after the charge $\delta Q$ was added to the
plate. The work done on $dQ$ by $q$ can be found by rewriting the
integral we have started with as
$$\int \limits_{Q_0}^Q = \int \limits_{Q_1}^Q - \int
\limits_{Q_1}^{Q_0},$$ where $Q_1$ is an arbitrary point at
infinity. If we apply the previous argument to both integrals on
the right hand side, we obtain
$$dA = - q [\Phi_{dQ}(P, Q) - \Phi_{dQ}(P, Q_0)].$$
The formulas for $dA$ can be summarized as
$$dA = -qd\Phi,$$
where $d\Phi$ is the contribution of the charge $dQ$ to the total
variation of the potential $\Phi$. Finally, we sum up all
contributions to the total work to obtain
\begin{equation}
\delta A = - q \delta \Phi. \label{eq:woe}
\end{equation}

\vskip 2mm With the formula for the work of a pointlike charge at
hand, it is straightforward to compute the work of a dipole (fig.
\ref{fig:eld} to the right). Consider an elementary dipole with
the moment $\bf d$, that is, a couple of pointlike charges $+ q$
and $- q$ displaced with respect to each other by the vector $\bf
a$ (so that ${\bf d} = q \bf a$), taken in the limit in which $q
\to \infty$, $a \to 0$ and $qa$ is finite. With the charge $+ q$
located at the point $P_+$ and the charge $- q$ located at the
point $P_-$, it holds
$$\delta A = - q \delta \Phi (P_+) + q \delta \Phi (P_-) = - q {\bf a}\
.\ \nabla \delta \Phi,$$ and after exchanging $\nabla$ and
$\delta$ and using the expression of $\bf d$ in terms of $q$ and
$\bf a$, we find
\begin{equation}
\delta A = {\bf d}\ .\  \delta \vec{\cal E}, \label{eq:woe1}
\end{equation}
where $\delta \vec{\cal E}$ is the variation of the electric
intensity $\vec{\cal E}$ at the point where the dipole is located.

\vskip 2mm The two expressions for the work we have arrived at are
consistent with the expressions for the potential energy of a
pointlike charge
\begin{equation}
E = q \Phi \label{eq:ene}
\end{equation}
and of a dipole
\begin{equation}
E = - {\bf d}\ .\ \vec{\cal E}. \label{eq:ene1}
\end{equation}
As can be seen by comparing these equations with equations
(\ref{eq:woe}) and (\ref{eq:woe1}), it holds $\delta A = - \delta
E$; thus, the work done equals the energy lost.

\vskip 2mm To gain a better insight into the relation between
$\delta A$ and $\delta E$, let us rederive it via the formula for
the energy of electrostatic field. Consider a static, spatially
bounded system of charges consisting of the subsystem $S$,
generating the electric field with the potential $\Phi$ and
intensity $\vec{\cal E}$, and the subsystem $s$, generating the
electric field with the potential $\phi$ and intensity $\bm
\epsilon$. The interaction energy of the two fields is
$$E_{int} = \epsilon_0 \int \vec{\cal E}\ .\ {\bm \epsilon}
dV,$$ and can be rewritten either as
$$E_{int} = - \epsilon_0 \int \nabla \Phi\ .\ {\bm
\epsilon} dV = \epsilon_0 \int \Phi \nabla \ .\ {\bm \epsilon} dV
= \int \Phi \rho^{(s)} dV,$$ or as
$$E_{int} = - \epsilon_0 \int \vec{\cal E}\ .\ \nabla \phi
dV = \epsilon_0 \int \phi \nabla \ .\ \vec{\cal E} dV = \int \phi
\rho^{(S)} dV,$$ where $\rho^{(s)}$ and $\rho^{(S)}$ are the
charge densities of the systems $s$ and $S$ respectively. After
passing to the discrete charges $Q_I$ (in the system $S$) and
$q_i$ (in the system $s$), we obtain
\begin{equation}
E_{int} = \sum q_i \Phi_i = \sum Q_I \phi_I, \label{eq:einte}
\end{equation}
where $\Phi_i$ is the value of $\Phi$ at the location of the
charge $q_i$ and $\phi_I$ is the value of $\phi$ at the location
of the charge $Q_I$. The first expression is the potential energy
$E$ of the system $s$ in the field of the system $S$, and the
second expression is the potential energy $E'$ of the system $S$
in the field of the system $s$. If the system $s$ is kept fixed,
like the pointlike charge and the dipole in the previous
discussion, and if the charge of the system $S$ is
redistributed--which may include bringing some charge from
infinity to the neighborhood of the system $s$--the resulting
variations of $E$ and $E'$ coincide. Since the latter variation
equals minus the work $\delta A$ done by the system $s$ on the
system $S$, we arrive again at the formula $\delta A = - \delta
E$.

\begin{figure}[ht]
\centerline{\includegraphics[width=12.5cm]{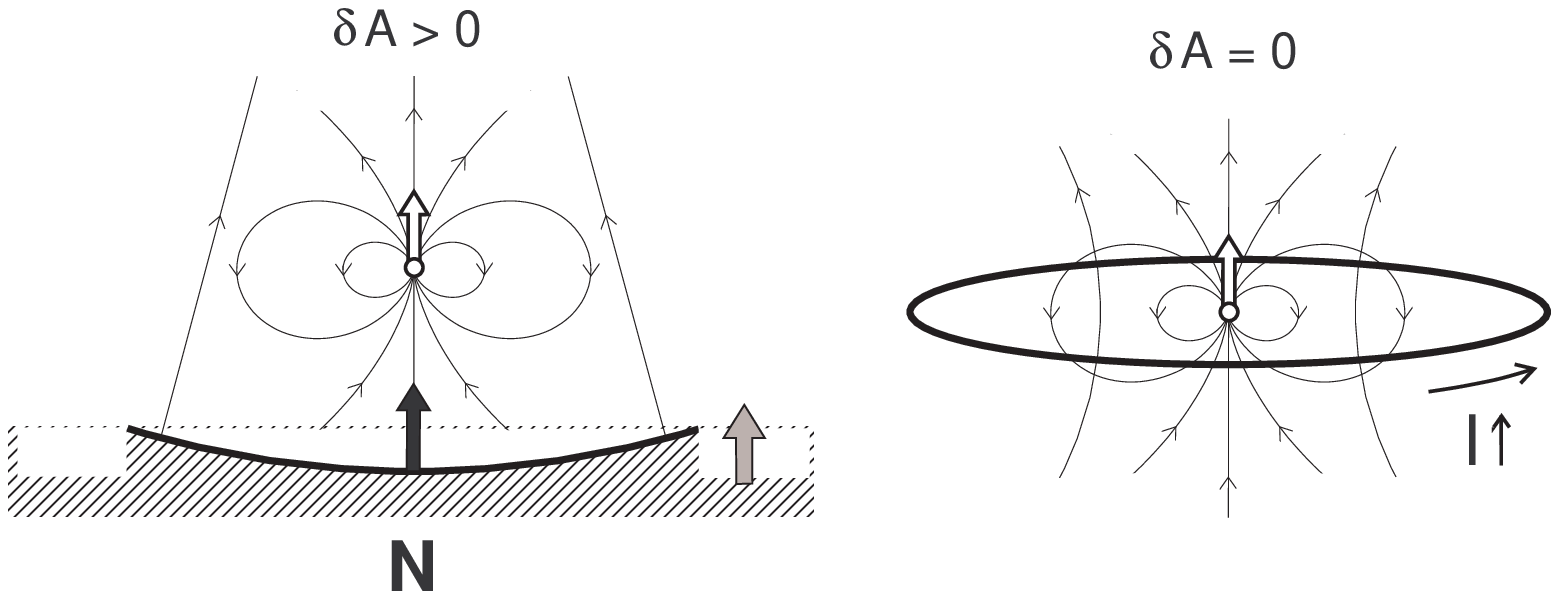}}
\caption{Spin in a varying magnetic field} \label{fig:magd}
\end{figure}

\vskip 2mm {\it 3. Magnetic dipole in a varying magnetic field.}
After having analyzed the effect of a varying electric field on a
pointlike charge and a dipole, let us turn to the effect of a
varying magnetic field on a spinning particle. The field can be
generated either by a moving magnet, or by a moving circuit, or by
a circuit with a varying current. Consider first the case with the
moving magnet. Let a spinning particle with the magnetic moment
$\bf m$ be placed nearby a permanent magnet generating an
inhomogeneous magnetic field with the induction $\bf B$ (fig.
\ref{fig:magd} to the left). After the magnet is shifted, the
magnetic field at the point where the particle is located changes
by $\delta \bf B$ and the particle does the work $\delta A$. The
relation
between these two quantities is most easily established with the
help of the electrostatic analogy, by replacing $\bf d$ by $\bf m$
and $\vec{\cal E}$ by $\bf B$ in equation (\ref{eq:woe1}). In this
way we obtain
\begin{equation}
\delta A = {\bf m}\ .\  \delta {\bf B}. \label{eq:wom}
\end{equation}
For a string of spins that is could or could not be in contact
with heat reservoir, this yields equation (\ref{eq:wo}). Note that
one can arrive at the expression (\ref{eq:wom}) for $\delta A$
also without the reference to electrostatics. The reverse process
to the one considered, with the spin moving and the magnet at
rest, takes place in the Stern-Gerlach experiment. If one replaces
the spinning particle by two nearby monopoles with opposite signs,
one immediately obtains the well-known formula for the force
acting on the particle,
$${\bf f}_{SG} = {\bf m}\ .\  \nabla {\bf B}.$$
By the law of action and reaction, the particle acts on the magnet
by the force ${\bf f} = - {\bf f}_{SG}$. If the magnet is moving
and the spin stays at rest, the work done on the magnet is
$$\delta A = {\bf f}\ .\  \delta {\bf r} = {\bf f}_{SG}\ .\
\delta {\bf r}',$$ where $\delta {\bf r}$ is the displacement of
the magnet and $\delta {\bf r}'$ is the displacement of the
particle with respect to the magnet; and if we insert here the
expression for ${\bf f}_{SG}$, we obtain again equation
(\ref{eq:wom}).

\vskip 2mm Let us now replace the magnet by the circuit. The
previous reasoning seems to stay valid as long as we restrict
ourselves to the circuits that move; for example, a circuit that
approaches the spin without varying its shape, or a circuit that
contracts towards the spin. Then, the Lorentz forces by which the
spin is acting on mobile charges inside the circuit do work on the
bulk of the circuit, and it turns out that the work has just the
right value,
\begin{equation}
\delta A_L = {\bf m}\ .\  \delta {\bf B}. \label{eq:womL}
\end{equation}
To check that this is the case, one needs three ingredients: the
formula for the Lorentz force, the Biot-Savart law and the
expression for the potential of the dipole field. We will spare
ourselves this calculation since, as we will see, it can be
carried out `in one line' with the help of the formula for the
interaction energy of the magnetic fields.

\begin{figure}[ht]
\centerline{\includegraphics[width=5cm]{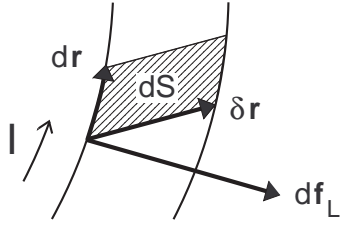}} \caption{Work
done by the Lorentz forces} \label{fig:wo}
\end{figure}

\vskip 2mm Equations (\ref{eq:wom}) and (\ref{eq:womL}) suggest
that the behavior of a magnetic dipole in a varying field is the
same as the behavior of an electric dipole in the same situation,
no matter how the field is generated. However, the whole concept
falls apart when the circuit neither moves as a solid body nor
deforms, and the magnetic field acting on the spin changes because
the current that flows through the circuit is changing (fig.
\ref{fig:magd} to the right). The Lorentz force is perpendicular
to the direction of motion of the charges in the circuit, hence it
does no work on them; and there is no other mechanism for doing
work in sight. This casts doubts also on the previous analysis
concerning the moving circuits. At a closer look we find that we
have omitted an important part of the story; namely, we paid no
attention to the fact that the particle generates an additional
voltage in the moving circuit according to the Faraday's law of
induction. The induced voltage does work on the mobile charges in
the circuit, or work must be done on the charges by the external
forces in order to overcome the induced voltage; and it is a
matter of simple calculation to check that this work exactly
cancels the work done by the Lorentz forces. Since this result is
important for further considerations, let us prove it here.
Consider an electric circuit with the current $I$ in the magnetic
field $\bf b$ of the spinning particle, and suppose an
infinitesimal segment of the circuit $d{\bf r}$ shifts by the
vector $\delta {\bf r}$, as depicted in fig. \ref{fig:wo}.
The Lorentz force acting on the segment is $d{\bf f}_L = Id{\bf
r}\ \times\ {\bf b}$, hence the work done on the segment is
$$dA_L = d{\bf f}_L\ .\ \delta {\bf r} = I{\bf b}\ .\ (\delta {\bf r}\
\times\ d{\bf r}) = I{\bf b}\ .\ d{\bf S} = Idf,$$ where $d{\bf
S}$ is the element of the surface spanned by the vectors $\delta
{\bf r}$ and $d{\bf r}$, and $df$ is the variation of the magnetic
flux through the circuit due to the displacement of the segment.
As a result, the total work done by the Lorentz forces is
$$\delta A_L = I\delta f,$$
where $\delta f$ is the total variation of the magnetic flux. On
the other hand, the work done on the circuit by the spinning
particle via electromagnetic induction is
$$\delta A_{ind} = I U_{ind} dt,$$
where $U_{ind}$ is the induced voltage in the circuit and $dt$ is
the duration of the process; and since, by the Faraday's law of
induction, it holds $U_{ind} = - df/dt$, we obtain
$$\delta A_{ind} = - I\delta f.$$
As stated above, the two works cancel,
\begin{equation}
\delta A_L + \delta A_{ind} = 0. \label{eq:wom1}
\end{equation}
Note that this can be derived `in one line', too, if one uses the
fact that the induced electric field in the moving circuit comes
from the relativistic transformation of the magnetic field into
the rest frames of the elements of the circuit. To summarize,
there is a sharp distinction between the case when the varying
magnetic field is generated by a permanent magnet, and the case
when its source is a circuit with an electric current. In the
former case, the work done by the spinning particle is nonzero and
given by the formula (\ref{eq:wom}), while in the latter case the
work is zero.

\vskip 2mm {\it 4. Energy of the magnetic dipole.} The conclusion
we have arrived at provokes a question about the energy balance.
For a magnetic dipole in an external magnetic field, one generally
uses the expression for energy analogical to that from
electrostatics,
\begin{equation}
E = - {\bf m}\ .\ {\bf B}. \label{eq:enm}
\end{equation}
The previous analysis leaves an impression that the energy of the
dipole equals sometimes this expression and sometimes zero. But
the matters are even more mixed up. It turns out that the total
energy of the dipole is actually given by equation (\ref{eq:enm})
with the {\it reversed sign},
\begin{equation}
E_{tot} = {\bf m}\ .\ {\bf B}. \label{eq:enm1}
\end{equation}
More precisely, this is true if the dipole as well as the source
of the external magnetic field are purely classical (non-quantum)
objects, since otherwise they cannot be described by the Maxwell
theory in a consistent way. The requirement is obeyed neither by
spinning particles like electrons or atoms, nor by permanent
magnets. As we will see, these objects can be still included into
the theory, but we must be cautious when doing that.

\vskip 2mm A nice derivation of $E_{tot}$, by computing the total
work needed to bring an electrical circuit into a magnetic field,
can be found in \S 15.2 of Feynman lectures. For comparison, in
Jackson's textbook the expression for $E_{tot}$ is not even
written down, not to mention derived. It is only suggested, in a
comment on equation (\ref{eq:enm}) (on p.~185 of the 1975
edition), that $E_{tot}$ differs from $E$ because `in bringing the
dipole $\bf m$ into its final position in the field, work must be
done to keep the current $\bf J$ which produces $\bf m$ constant'.
However, as the reader of Feynman knows, this is not the whole
story, since work must be done also to keep the current which
produces the {\it external field} constant. After computing the
potential energy of a magnetic medium put into a magnetic field,
Jackson comments on $E_{tot}$ again (on p.~217 of the 1975
edition), but in a rather cryptic way and as before without giving
the explicit expression for it. On the contrary, the Feynman's
derivation of $E_{tot}$ is completely satisfactory and we could
conclude the discussion just by giving the reference to it.
However, for our purposes it is useful to compute $E_{tot}$ in a
different way, as the interaction energy of fields.

\vskip 2mm Consider an elementary magnetic dipole with the moment
$\bf m$, that is, a circuit with the current $i$ and the vector of
the surface $\bm \sigma$ (so that ${\bf m} = i{\bm \sigma}$),
taken in the limit in which $i \to \infty$, $\sigma \to 0$ and
$i\sigma$ is finite. Let the dipole be put into a magnetic field
generated by a system of stationary, spatially bounded currents,
and denote the magnetic fields of the dipole and the currents by
$\bf b$ and $\bf B$ respectively. We will show that the
interaction energy of the two fields,
\begin{equation}
E_{int} = \frac 1{\mu_0} \int {\bf B}\ .\ {\bf b} dV,
\label{eq:eintdf}
\end{equation}
equals just the energy $E_{tot}$ from equation (\ref{eq:enm1}).
Denote  the distance between the dipole and the observer by $r$
and the unit vector pointing from the dipole to the observer by
$\bf n$. Our starting point will be the formula for the vector
potential of the dipole
$${\bf a} = \frac {\mu_0}{4\pi}\ \frac {{\bf m}\ \times\
{\bf n}}{r^2}.$$ For {\it finite} circuits, this holds only at
large distances, but since we have shrunk the circuit to a point,
we can write $\bf a$ in this way in the whole space. The magnetic
field of the dipole is
$${\bf b} = \nabla\ \times\ {\bf a} = -\frac {\mu_0}{4\pi} \nabla\
\times\ \left({\bf m}\ \times\ \nabla \frac 1r\right) = \frac
{\mu_0}{4\pi} \left[({\bf m}\ .\ \nabla) \nabla \frac 1r - {\bf
m}\triangle \frac 1r\right] = - \nabla \phi + \mu_0 {\bf m} \delta
({\bf r}),$$ where $\phi$ is the potential of the field,
$$\phi = \frac {\mu_0}{4\pi}\ \frac {{\bf m}\ .\ {\bf n}}{r^2},$$
and $\bf r$ is the radius vector of the observer with respect to
the point where the dipole is located. After inserting this into
$E_{int}$, we find
$$E_{int} = - \frac 1{\mu_0} \int {\bf B}\ .\ \nabla \phi dV +
{\bf m}\ .\ {\bf B} = \frac 1{\mu_0} \int \phi \nabla\ .\ {\bf B}
dV + {\bf m}\ .\ {\bf B} = {\bf m}\ .\ {\bf B},$$ q. e. d.

\vskip 2mm We have argued that the dipole, disguised under the
name `spinning particle', does zero work in a varying magnetic
field if the field is generated by electric currents. This is
consistent with the fact that the dipole has a nonzero total
energy which varies in the process, since the sources of the field
do work on the dipole via electromagnetic induction. We have just
to show that this work equals minus the variation of the energy of
the dipole. To carry out the task, let us perform analogical
manipulations with the formula for $E_{int}$ as in the
electrostatic case. Consider a stationary, spatially bounded
system of currents consisting of the subsystem $S$, generating the
magnetic field with the potential $\bf A$ and intensity $\bf B$,
and the subsystem $s$, generating the magnetic field with the
potential $\bf a$ and intensity $\bf b$. The interaction energy of
the two fields, defined in equation (\ref{eq:eintdf}), can be
rewritten as
$$E_{int} = \frac 1{\mu_0} \int (\nabla\ \times\ {\bf A})\ .\ {\bf
b} dV = \frac 1{\mu_0} \int {\bf A}\ .\ (\nabla\ \times {\bf b})
dV = \int {\bf A} \ .\ {\bf j}^{(s)} dV,$$ or as
$$E_{int} = \frac 1{\mu_0} \int {\bf B}\ .\ (\nabla\ \times\ {\bf
a}) dV = \frac 1{\mu_0} \int {\bf a}\ .\ (\nabla\ \times {\bf B})
dV = \int {\bf a} \ .\ {\bf j}^{(S)} dV,$$ where ${\bf j}^{(s)}$
and ${\bf j}^{(S)}$ are the current densities of the systems $s$
and $S$ respectively. After passing to the circuits $C_A$ with the
currents $I_A$ (in the system $S$) and the circuits $c_a$ with the
currents $i_a$ (in the system $s$), and denoting the surfaces
spanned by the circuits $C_A$ and $c_a$ by $S_A$ and $s_a$, we
obtain
$$E_{int} = \sum i_a \oint \limits_{c_a} {\bf A} \ .\ d{\bf r} =
\sum i_a \int \limits_{s_a} {\bf B} \ .\ d{\bf S}$$ and
$$E_{int} = \sum I_A \oint \limits_{C_A} {\bf a} \ .\ d{\bf r} =
\sum I_A \int \limits_{S_A} {\bf b} \ .\ d{\bf S}.$$ In the
compact notation,
\begin{equation}
E_{int} = \sum i_a F_a = \sum I_A f_A, \label{eq:eintm}
\end{equation}
where $F_a$ is the flux of $\bf B$ through the surface $s_a$ and
$f_A$ is the flux of $\bf b$ through the surface $S_A$. The first
expression can be identified with the total energy $E_{tot}$ of
the system $s$ in the field of the system $S$, and the second
expression can be regarded as the total energy $E_{tot}'$ of the
system $S$ in the field of the system $s$. If the system $s$
consists of one elementary dipole, the first expression yields
$$E_{int} = iF = i {\bm \sigma}\ .\ {\bf B} = {\bf m} \ .\ {\bf
B},$$ in accordance with the more direct calculation carried out
above. Suppose we leave the system $s$ unchanged and change the
currents, the location of the circuits and/or the shape of the
circuits in the system $S$. Then, the energy of the system $s$
changes by
$$dE_{tot} = \sum i_a dF_a = - \sum i_a u_{ind,a} dt = - \delta
a_{ind},$$ where $u_{ind,a}$ is the induced voltage in the circuit
$c_a$, $dt$ is the duration of the process and $\delta a_{ind}$ is
the total work of the voltages $u_{ind,a}$. The energy of the
system $S$ changes by the same amount, $dE_{tot}' = dE_{tot}$. To
have the energy budget balanced, we must assume that the work of
the induced voltage is done by the external sources on the circuit
in which the voltage appears. Then, the system $S$ does the work
$\delta a_{ind}$ on the system $s$, while the system $s$ does no
net work on the system $S$. Indeed, if the currents are varying
only, no work is done at all; and if the circuits move, the work
done by the Lorentz forces is compensated by the work done by the
induced voltages, see equation (\ref{eq:wom1}). Thus, the energies
$E_{tot}$ and $E_{tot}'$ do not change because the system $s$ does
work on the system $S$, as in the electrostatic case, but because
the system $S$ does work on the system $s$. This scheme has been
outlined in the remark about the behavior of a dipole in a varying
field we have started with; and the condition of consistency
mentioned there is just the relation between $dE_{tot}$ and
$\delta a_{ind}$.

\vskip 2mm To complete the analysis, let us define the {\it
mechanical} energy of the system $s$ in the field of the system
$S$ as $E = - E_{tot}$. Then, by mimicking the procedure by which
we have obtained the relation between $dE_{tot}$ and $\delta
a_{ind}$, we can obtain a relation between $dE'$ (the variation of
the mechanical energy of the system $S$ in the field of the system
$s$) and $\delta A_{ind}$ (the work done on the system $S$ by the
system $s$ via electromagnetic induction),
$$dE' = \delta A_{ind}.$$
Since $\delta A_L = - \delta A_{ind}$, we can also write
$$dE' = - \delta A_L,$$
which explains the term `mechanical energy' for $E$. Furthermore,
from $dE = dE'$ it follows $dE = - \delta A_L$; and by utilizing
the expression (\ref{eq:enm}) for $E$, we find the expression
(\ref{eq:womL}) for $\delta A_L$. Finally, let us consider an
elementary dipole in a {\it constant} magnetic field, and discuss
the variation of its energy due to the change of its orientation.
To keep the notations unchanged, let us identify, as before, the
dipole with the system $s$, and exchange the roles of the systems
$s$ and $S$. In this way we obtain
$$dE_{tot} = - \delta A_{ind} = - \delta a_L - \delta a_{ind} -
\delta A_{ind},$$ where the only new quantity is the work $\delta
a_L$ done on the dipole by the external field via the Lorentz
forces. If we insert here from $dE = - \delta a_L$, we have
\begin{equation}
dE_{tot} = dE - \delta a_{ind} - \delta A_{ind}. \label{eq:etot}
\end{equation}
The three terms on the right hand side correspond to the three
contributions to $E_{tot}$ in Feynman's \S 15.2. Both $\delta
a_{ind}$ and $\delta A_{ind}$ equal $dE$, hence it holds $dE_{tot}
= - dE$ (as it should, considering how we have defined $E$).

\vskip 2mm {\it 5. Role of quantum mechanics.} Thermodynamics of a
string of spins was studied in experiments with negative
temperatures. The experiments used a crystal put into a strong
magnetic field, with the original magnetization reversed by a
discharging condenser. In principle, one could obtain the effect
we are interested in by adding a permanent magnet to the setup,
and moving it towards the crystal or away from it. We will not
discuss how this could be done in practice; our aim is just to
specify what kind of spinning particles and sources of magnetic
field can be used to extract work from the field.

\vskip 2mm In the actual experiment, spinning particles were
nuclei in the crystal, and in the extended version proposed here,
the source would a ferromagnetic material or, in a higher
resolution, ions in the domains of which the material consists. In
both cases we are dealing with {\it quantum-mechanical} particles,
therefore the previous reasoning about the energy balance, based
on the {\it classical} Maxwell theory, must be revised. The
particles are composite, so that they can occupy different energy
levels in their rest frame; however, in the setting we are
considering they remain all the time in the ground state. If such
particle is put into a varying field or moves in a stationary
field, the work $\delta a_{ind}$ is {\it not} to be included into
its energy balance. The reason is obvious: even if there are
electric currents inside the particle, the electromagnetic
induction has no effect on them. The situation is effectively the
same as with elementary particles as electron or (in the given
context) nucleon. In the Maxwell theory, both kinds of particles
can be modeled as tiny self-sustained electrical circuits with no
response to electromagnetic induction. Elimination of the work
$\delta a_{ind}$ for ions in the moving magnet leads to
elimination of the work $\delta A_{ind}$ done on the magnet by the
spinning particle staying at rest; and with both works $\delta
a_{ind}$ and $\delta A_{ind}$ eliminated, the total energy gain
from the shift of the magnet with respect to the particle equals,
as desired, the work $\delta A$ of equation (\ref{eq:wom}).

\vskip 2mm For composite particles that {\it can} be excited, one
can argue that $\delta a_{ind}$ is still irrelevant if they just
change their orientation in the external field, but can be
relevant if they shift with respect to field, or stay at rest and
the field varies. For the problem with a particle changing its
orientation, consider how the Zeeman effect is described in QED
(see, for example, Landau-Lifshitz IV, \S 51). In the presence of
a magnetic field, the energy levels $E_{at}$ of the atom are
modified by the mechanical part of the energy $E_{tot}$ only, and
the energy of the emitted or absorbed photon is given by $\Delta
E_{at}$, with no contribution of $\delta a_{ind}$. This is
presumably the consequence of the fact that we restrict ourselves
to the lowest order of perturbation theory, since then the effects
of the interaction of the atom with the magnetic field and with
the photon just sum up, with no interference between them.
However, it seems strange to use such approximation if we know
from the Maxwell theory that the work $\delta a_{ind}$ is not much
less in the absolute value than the energy difference $\Delta E$,
but equal to it. To remove the apparent inconsistency, let us
observe that $\delta a_{ind}$ can be viewed as the energy spared
by the battery that feeds the circuit, in case the induced voltage
appears in the circuit and replaces a part of the emf of the
battery. The energy can be turned into work, say, by using the
residual emf to drive an electric motor. For a quantum-mechanical
system like atom, an analogue of the battery would be an
energy-supplying device taking the system back into the stationary
state it occupied before, anytime it leaves it into a state with a
different magnitude of magnetic moment because of spontaneous
emission. Since the processes of absorbtion and emission of
photons due to the Zeeman effect are well described as transitions
between stationary states, the `battery' is not participating in
them and the work $\delta a_{ind}$ does not show up. A different
question is whether the `battery' should not be put into action if
the system shifts in a stationary magnetic field or is located at
a fixed place in a varying magnetic field. Then, the relaxation
times for the transitions between the stationary states of the
system are to be included into considerations.

\vskip 2mm The description of both elementary and composed
particles in the Maxwell theory as self-sustained circuits seemed
to work well in the previous analysis. However, if one regards
electrons as {\it truly} elementary, one can ask whether they
should not be represented better as `electric-like' magnetic
dipoles; that is, as pairs of magnetic monopoles with opposite
signs placed close to each other. We mentioned this representation
when motivating the formula for the Stern-Gerlach force. The two
kinds of elementary dipole, the pair of monopoles and the
electrical circuit, are in almost all respects undistinguishable.
However, the question about the correct way how to represent
electrons can still be decided experimentally; and as we will see,
the answer is `by circuits'.

\vskip 2mm Magnetic fields of the two kinds of dipole differ only
by the value they assume, in the sense of the theory of
distributions, at the point where the dipole is located. If we
denote the magnetic field of the monopole-based dipole by ${\bf
b}_I$ and the magnetic field of the circuit-based dipole by ${\bf
b}_{II}$, we have
$${\bf b}_I = - \nabla \phi,\ \ {\bf b}_{II} = - \nabla \phi +
\mu_0 {\bf m} \delta ({\bf r}).$$ To be able to compare the
physical effects of the two fields, we must split the vector field
$\nabla \phi$, with $\phi \propto {\bf m}\ .\ {\bf n}/r^2$, into a
regular part and a part proportional to the $\delta$-function. (A
similar splitting of $\nabla\ \times\ \bf a$ is carried out in
Jackson; however, the procedure presented there is restricted to
the first step of our procedure.) First, consider the integral of
$\nabla \phi$ over an arbitrary ball with the center at the
origin, where the dipole is located. If we attempted to compute
the integral directly, we would obtain an ill defined expression
of the form `infinity times zero', where the infinity comes from
the integration over $r$ and zero comes from the integration over
the angles. However, we can evaluate the integral--in fact, {\it
define} it--by using the Gauss theorem, in a similar way as we
evaluate the integral of the function $\triangle(1/r)$ over an
arbitrary domain containing the origin. This yields
$$\int \nabla \phi dV = \oint \phi d{\bf S} = \frac {\mu_0}{4\pi}
\oint \frac {({\bf m}\ .\ {\bf n}) d{\bf S}}{r^2} = \mu_0 \big
\langle ({\bf m}\ .\ {\bf n}) {\bf n} \big \rangle = \frac 13
\mu_0 {\bf m},$$ where the angle brackets in the last but one term
denote averaging over angles. Consider now a smooth spherically
symmetric function $f(r)$, decreasing at infinity at least as
$r^{-p}$ with some $p > 0$, and compute the integral of the
product $f\nabla \phi$ over the whole space. Let us divide the
integration domain into two parts: the ball ${\cal B}_\epsilon$,
with the center at the origin and the radius $\epsilon$ such that
the Taylor expansion of $f$ at the origin converges uniformly in
${\cal B}_\epsilon$, and the rest of the space. The integral over
the rest of the space, as well as the integrals of the higher
order terms of the Taylor expansion in ${\cal B}_\epsilon$, are
zero because the integral over the angles is zero and the
integrals over $r$ are finite. Thus
$$\int f \nabla \phi dV = \int \limits_{{\cal B}_\epsilon} f\nabla
\phi dV = f(0) \int \limits_{{\cal B}_\epsilon} \nabla \phi dV +
\frac 12 f''(0) \int \limits_{{\cal B}_\epsilon} r^2 \nabla \phi
dV + \ldots = \frac 13 \mu_0 {\bf m} f(0).$$ This suggests that we
can write the vector field $\nabla \phi$ as
$$\nabla \phi = \nabla \phi_{reg} + \frac 13 \mu_0 {\bf m} \delta
({\bf r}),$$ with a properly defined $\nabla \phi_{reg}$. The
definition reads
$$f\nabla \phi_{reg} = [f - f_0(r)] \nabla \phi,$$
where $f = f({\bf r})$ is an {\it arbitrary} function of $\bf r$
and $f_0 (r)$ is a spherically symmetric function equal to $f(0)$
at the origin and decreasing at least as $r^{-p}$ at infinity. (By
subtracting $f_0$ from $f$ we have regularized $f$ to zero at the
origin in such a way that $\nabla \phi$ weighted by the
regularized function is integrable.) With the $\delta$-function
term separated out from $\nabla \phi$, we can write the two
versions of the dipole magnetic fields as
\begin{equation}
{\bf b}_I = - \nabla \phi_{reg} - \frac 13 \mu_0 {\bf m} \delta
({\bf r}),\ \ {\bf b}_{II} = - \nabla \phi_{reg} + \frac 23 \mu_0
{\bf m} \delta ({\bf r}). \label{eq:fields}
\end{equation}
The term in the expression for the magnetic field of the electron
proportional to the $\delta$-function makes all the difference
when one computes the hyperfine splitting of the ground state of
the hydrogen atom. The splitting produces the 21-cm hydrogen line,
famous for its applications in astrophysics. By performing the
computation with the fields ${\bf b}_I$ and ${\bf b}_{II}$, we
obtain the wavelength of the transition between the split energy
levels 42~cm and 21~cm, respectively. Thus, there is a strong
experimental evidence that the correct choice for the magnetic
field of electron is ${\bf b}_{II}$.

\vskip 2mm {\it Acknowledgement.} I am grateful to Vladim\'ir \v
Cern\'y for friendly arguments about the ideas explained here.

\vskip 6mm \noindent {\Large \bf References}

\vskip 2mm \noindent Feynman R., Leighton R., Sands M.: {\it The
Feynman Lectures on Physics}, Adison Wesley (1970).

\vskip 2mm \noindent Jackson J. D.: {\it Classical
Electrodynamics}, John Wiley \& Sons (1975).

\vskip 2mm \noindent Berestetskii V. B., Lifshitz E. M.,
Pitaevskii L. P.: {\it Quantum Electrodynamics},
Butterworth-Heine\-mann (1982).

\enddocument